\documentclass[12pt,a4paper]{article}
\usepackage{natbib}
\usepackage{array,lscape}

\begin{document}
\title{Computational Modelling of Plasticity-Led Evolution}
\author{Eden Tian Hwa Ng and Akira R. Kinjo\\
  Department of Mathematics, Faculty of Science, \\
  Universiti Brunei Darussalam}

\maketitle
\abstract{%
Plasticity-led evolution is a form of evolution where a change in the environment induces novel traits via phenotypic plasticity, after which the novel traits are genetically accommodated over generations under the novel environment.  This mode of evolution is expected to resolve the problem of gradualism (i.e., evolution by the slow accumulation of mutations that induce phenotypic variation) implied by the Modern Evolutionary Synthesis, in the face of a large environmental change. While experimental works are essential for validating that plasticity-led evolution indeed happened, we need computational models to gain insight into its underlying mechanisms and make qualitative predictions. Such computational models should include the developmental process and gene-environment interactions in addition to genetics and natural selection. We point out that gene regulatory network models can incorporate all the above notions. In this review, we highlight results from computational modelling of gene regulatory networks that consolidate the criteria of plasticity-led evolution. Since gene regulatory networks are mathematically equivalent to artificial recurrent neural networks, we also discuss their analogies and discrepancies, which may help further understand the mechanisms underlying plasticity-led evolution.
}


  \section{Introduction}\label{sec1}

Integrating natural selection and genetics, the Modern Evolutionary Synthesis has been the cornerstone of evolutionary models since its inception in the early 20th century \citep{smith1998evolutionary, nowak2006evolutionary, broom2013game}. However, its gene-centric nature has been criticized: under Modern Evolutionary Synthesis, all possible variation in the expressed phenotypes of individuals of a species can only be explained in terms of genetic mutations \citep{laland2014does}. Conventional quantitative genetics models can incorporate gene-environment interactions (i.e., $G\times E$ covariance), but they either ignore such interactions \citep{CrowANDKimura} or treat them as part of genetic components (because the specific response to specific environmental cues is considered genetically encoded) \citep{falconer1996quantitative}. Hence, the only means for an individual to survive a large environmental change is to possess mutations that produce a phenotype that is already adapted to the novel environment. Given that mutations are random and selection favors phenotypes that are adapted to the current environment, the preexistence of adaptive phenotypes to future novel environments seems extremely unlikely.



One possible resolution for the above paradox is to consider phenotypic changes in response to environmental changes without mutations \citep{west2003developmental}. This allows possibility for individuals to express different phenotypes in different environments, potentially increasing fitness under a novel environment. Moreover, while possession of an adaptive mutation only affects few individuals in the population, an environmental change affects the entire population. In fact, this environment-sensitive variation in traits is called phenotypic plasticity and is ubiquitous in nature \citep{west2003developmental, gilbert2009ecological}. Examples of phenotypic plasticity include, but are not limited to, temperature dependent sex determination in reptiles \citep{whiteley2021two}, behavioral and physical dimorphism in dung beetles \citep{emlen1997alternative} and caste polyphenism in social insect populations \citep{weiner2012epigenetics}. Phenotypic plasticity arises from the developmental process through which genetic and environmental information are integrated \citep{gilbert2009ecological, merila2004variation, nishikawa2018mechanism, schulz2010dutch, schneider2014regulatory, west2003developmental}. Hence, plastic traits are partially acquired traits. Since plastic traits are not heritable, the Modern Evolutionary Synthesis has largely neglected the developmental process that produces them \citep{muller2007evo, laland2014does}. However, there is accumulating evidence that demonstrates the role of phenotypic plasticity in evolution \citep{west2003developmental, ehrenreich2016genetic,  scheiner2021loss, pfennig2021phenotypic}.

A classical example is Waddington's observation of genetic assimilation in fruit flies \citep{waddington1953genetic}. The \emph{crossveinless} wing pattern which was initially only expressed under heat shock becomes expressed regardless of exposure to heat shock or not after some 20 generations of selective breeding, a phenomenon which Waddington called genetic assimilation. Waddington's experiment shows that adaptive evolution can start through environmental induction of novel traits and that formerly acquired traits can become heritable. Genetic assimilation was later generalized as genetic accommodation which refers to any adaptive \emph{genetic} change in the environmental regulation of a phenotype \citep{west2003developmental, ehrenreich2016genetic, scheiner2021loss}. We emphasize that even though a plastic trait does not require mutations to appear, genetic accommodation requires genetic change (through mutation or recombination) for adjusting the developmental mechanism of the plastic trait through evolution.




Since phenotypic plasticity produces a change in phenotype in response to a change in environment before mutations change the regulatory interactions in the genome, this describes a form of evolution called \emph{plasticity-led evolution}. Plasticity-led evolution as defined by \citet{levis2016evaluating} has four core criteria, the first two describe the response of the population when initially exposed to a novel environment and the latter two describe the evolution of expressed phenotypes of the population as the population adapts to the novel environment. These criteria are the following:
\begin{enumerate}
\item Novel adaptive traits are initially conditionally expressed, 
\item Cryptic genetic variation will be uncovered when populations are exposed to the novel environment, 
\item Novel adaptive traits undergo a change in environmental regulation,
\item Novel adaptive traits undergo adaptive refinement.
\end{enumerate}
{It is critical to distinguish between the plastic response per se and cryptic genetic mutations uncovered due to the plastic response.}
We note that the variation in plastic response to the environment may be and often is heritable. However, evolution by the selection against such a variation that exhibits adaptive plastic response should be considered mutation-led evolution, and it does not solve the problem of the conventional theory.
Plasticity-led evolution, in contrast, assumes that adaptive plastic response is a universal feature of developmental systems independent of particular mutations  \citep{west2003developmental, kovaka2020underdetermination}. To the best of our knowledge, this crucial assumption is yet to be validated.
{Regarding the universality of plastic response independent of particular mutations, \citet{bergman2003evolutionary} have demonstrated that evolutionary capacitance, a breakdown of which may be regarded as a plastic response, is a general feature of GRNs. Computational results suggest that the universal plastic response is also adaptive (see below), but some experimental results suggest otherwise \citep{Ghalambor2015}.}

 In this review, we will examine computational models in light of these criteria of plasticity-led evolution. \citet{kovaka2020underdetermination} has called for a ``middle-range theory'' to identify key signatures of plasticity-led evolution. Computational modelling is expected to form an important component in constructing such a middle-range theory.

\section{Computational modelling of evolution}




In essence, traditional computational models based on the Modern Evolutionary Synthesis incorporate the following notions \citep{laland2014does}:
\begin{enumerate}
\item Mutations produce variations in phenotypes.
\item Individuals expressing adaptive phenotypes are naturally selected.
\item Genotypes of selected individuals are passed on to their offsprings.
\end{enumerate}
More recently, there have been efforts to incorporate the effects of phenotypic plasticity into { quantitative genetics} models \citep{falconer1996quantitative, chevin2010adaptive, hangartner2022sexual, lande2009adaptation, nishikawa2014cooperation, rago2019adaptive, scheiner2020genetics} {(for earlier models, see \citet{scheiner1993genetics})}. { These models all neglect the developmental process. Primarily due to this limitation, phenotypic plasticity is necessarily treated as a genetically encoded variation. One notable exception is \citet{nishikawa2014cooperation} where phenotypic plasticity was treated as a random variable independent of genetic components. Therefore,} we seek a computational framework to integrate the developmental process with evolution.


A minimal model for investigating the evolution of developmental systems is the Wagner model \citep{wagner1996does}. The ingenuity of the Wagner model is to interpret the developmental process of an individual as the sequence of gene expression patterns which are determined by the following recursive equation:
\begin{equation}\label{eq:dev}
g_i(s+1) = \sigma\left(\sum_{j=1}^n G_{ij}g_j(s)\right)
\end{equation}
where  $g_i(s)$ is the gene expression level of the $i$-th gene at the $s$-th stage of development, $G_{ij}$ represents the regulatory effect of the $j$-th gene expression on the $i$-th gene expression and $\sigma$ is the step activation function. The matrix $G$ defines the gene regulatory network (GRN) of an individual and is interpreted as the genome. The steady state of the recursive equation is regarded as the adult phenotype which is subject to selection. Selected individuals are then paired for reproduction. Reproduction is modelled by swapping rows of the $G$ matrices of paired individuals. Genetic mutations are then introduced by randomly changing elements of the $G$ matrices. In some texts, developmental environmental cues are modelled as a perturbation to the initial gene expression pattern \citep{espinosa2011phenotypicrobustness, watson2014evolution}, while others model environmental cues as a developmental threshold \citep{draghi2012phenotypic, nagata2020emergence, kaneko2022evolution}. In particular, \citet{espinosa2011phenotypicrobustness} ingeniously modelled the environment as an inducer-selector pair of ``initial condition''-``optimal gene expression'' pair where each vector in the pair is allowed to vary independently from one another.

The Wagner model and its variants have been employed to investigate evolution of robustness \citep{wagner1996does, espinosa2011phenotypicrobustness, kaneko2022evolution}, evolutionary capacitance \citep{bergman2003evolutionary}, genetic assimilation \citep{masel2004genetic}, { developmental bias \citep{draghi2012phenotypic}}, relationship between robustness and non-genetic perturbations \citep{espinosa2011phenotypicrobustness}, evolution of novel phenotypes \citep{espinosa2011phenotypicplasticity, kaneko2022evolution}, evolution of modularity \citep{espinosa2010specialization}, selection for non-stationarity \citep{espinosa2016selection}, emergence of bistability \citep{nagata2020emergence} {and how phenotypic plasticity impacts evolution \citep{espinosa2021recombination, fierst2011history}}. We briefly summarized the key results { alongside their limitations} of these works in Table \ref{table: GRN}. { We stress that these limitations are not shortcomings of their models, but rather due to the specific goals of their works.} As can be seen in these works, the Wagner model seamlessly incorporates the biological principles of natural selection and genetics as well as the developmental process which interweaves genetic and environmental information into the phenotype \citep{chevin2022using}. By observing differences in adult phenotypes in response to different environmental cues, we see that the Wagner model can exhibit phenotypic plasticity through the developmental process. Due to these advantages over traditional evolutionary models, we believe that the Wagner model is a strong candidate for studying plasticity-led evolution. We will now discuss how well the Wagner model and its variants describe plasticity-led evolution.

\begin{landscape}
\begin{table}
\caption{\label{table: GRN} Gene Regulatory Network Models and their results}
\begin{tabular}{|m{5cm}|m{5cm}|m{5cm}|m{4.5cm}|}
\hline
Key features & Goals & Limitations & Reference\\
\hline
Different mechanisms of mutation, measure of robustness. & Evolution of plasticity. & No environment. &\cite{wagner1996does}\\
\hline 
Network pruning as analogy to knockout experiments. & Evolutionary capacitance of GRNs. & No environment &\cite{bergman2003evolutionary}\\
\hline
Environmental cue as threshold & Genetic assimilation in absence of selection & No environment-as-selector &\cite{masel2004genetic}\\
\hline
Environment cue as initial condition. & Accessibility of novel constitutive phenotype. & Environmental cue not integrated over development. &\cite{espinosa2011phenotypicplasticity}\\
\hline
Genotypic diversity, mutational and environmental robustness. & Relationship between robustness and phenotype variation in large environmental change & No correlation between environment-as-inducer and environment-as-selector. &\cite{espinosa2011phenotypicrobustness}\\
\hline
Direction of selection, evolvability & Developmental bias & Restrictive phenotype & \cite{draghi2012phenotypic}\\
\hline
Population average genotype, output prediction to novel input. & Link evolution and learning theory. & Adaptiveness of generalization & \cite{watson2014evolution}\\
\hline 
Selection for non-stationary phenotype, sexual populations & Robustness under selection for non-stationary phenotype & {Gene expression as phenotype} & \cite{espinosa2016selection}\\
\hline
Zero or unit environmental cue & Emergence and robustness of bistability & { Selection for difference in phenotypes instead for fitness} & \cite{nagata2020emergence} \\
\hline
Gene dependent regulatory threshold, multiple selected phenotypic optima & Effects of recombination on genetic accommodation & {No correlation between environment-as-inducer and environment-as-selector} & \cite{espinosa2021recombination}\\
\hline
\end{tabular}
\end{table}
\end{landscape}


\section{Plausibility of Plasticity-led Evolution}

Here, we compare the results from the Wagner model and its variants with each criterion given by \citet{levis2016evaluating}. By doing so, we aim to demonstrate the extent to which the Wagner model can explain plasticity-led evolution. We also aim to point out necessary innovations for future computational models of plasticity-led evolution (see Table \ref{table: PLE}). 

\begin{landscape}
\begin{table}
\centering
\caption{\label{table: PLE} GRN models (partly) supporting Plasticity-Led Evolution.}
\begin{tabular}{|m{2cm}|m{4cm}|m{4cm}|m{4cm}|m{3.5cm}|}
\hline
Criterion & Relevant notions & Existing Results & Prospective work & References\\
\hline
Adaptive plastic response in novel environments & Environment-as-inducer, Environment-as-selector, Plastic Response, Robustness  & Emergence of novel phenotype in novel environments. & Correlation between environment-as-inducer and environment-as-selector. & \cite{watson2014evolution} \\
\hline
Uncovering of cryptic mutations under large environmental change & Environment-as-inducer, Plastic Response, Robustness & Evolutionary capacitance and expression of cryptic mutations under stress. & Breakdown point of robustness. & \cite{bergman2003evolutionary, espinosa2011phenotypicrobustness}\\
\hline
Change in regulation or form & (Change in) Robustness, Environment-as-selector & Change in robustness alongside increased fitness over evolution. & Change in GRN structures accompanying adaptation. & \cite{wagner1996does, nagata2020emergence}\\
\hline
Adaptive refinement & Fitness, Environment-as-selector & Evolutionary algorithms trivially increase fitness & Refinement of GRN parameters. & \cite{wagner1996does, espinosa2011phenotypicplasticity}\\
\hline
\end{tabular}
\end{table}
\end{landscape}

\subsection{Conditional expression of novel adaptive traits}

Under plasticity-led evolution, when a population is first brought into a novel environment, adaptive traits should be expressed via plastic response \citep{west2003developmental, levis2016evaluating, levis2021innovation}. In other words, a change in environment induces a change in phenotype, which is more adaptive in the new environment. This prompts us to distinguish the two roles of the environment: ``inducer'' and ``selector''. The environment as an ``inducer'' gives rise to a response in phenotype. The environment as a ``selector'' discriminates between well adapted and poorly adapted phenotypes \citep{west1989phenotypic, espinosa2011phenotypicrobustness}. While individuals should exhibit a plastic response to large environmental changes, they should also have robustness against small environmental noise.


A computational model for studying adaptive plastic response should incorporate all the above features: environment as inducer and selector, plastic response and robustness. While there are many works investigating evolution of GRNs, none have explicitly evaluated models in light of adaptive plastic response. Hence, existing works study some of the features of plastic response, but none include all of them \citep{espinosa2011phenotypicrobustness, espinosa2016selection, nagata2020emergence, kaneko2022evolution}. 

One notable exception is \citet{watson2014evolution} who investigated the evolution of ``developmental memory'' using a variant of the Wagner model. They treated the environment-as-inducer as the initial condition and the environment-as-selector as the target phenotype. Their model was able to express multiple (trained) phenotypes given distinct inputs, which can be interpreted as phenotypic plasticity. The expressed phenotypes were shown to be robust to noise in the input.

We remark that the approach taken by \citet{watson2014evolution} which stresses on the correlation between the input and output of the GRN gives an insight to studying \emph{adaptive} plastic response. Here, a perturbation to the input induces an informed change to the output and yet this aspect is missing in most other works. By further enforcing correlation between the environment-as-selector and the environment-as-inducer, this allows one to determine whether such plastic response is adaptive or not.

\subsection{Uncovering of cryptic genetic variation}


In the context of plasticity-led evolution, mutations are said to be cryptic if their regulatory effects are not visible in an adapted environment but are seen under a novel environment. Such cryptic mutations accumulate under an adapted environment because they do not impact the expressed phenotype. On the other hand, under a novel environment, these formerly invisible mutations are expressed, resulting in increased variation in expressed phenotypes. This ability to harbor cryptic mutations for release under a large environmental change is called evolutionary capacitance and has been well regarded for playing a critical role for facilitating evolution \citep{rutherford1998hsp90, masel2005evolutionary, masel2013q, ehrenreich2016genetic}.



A computational model for studying evolutionary capacitance should exhibit the mutational robustness of phenotypes under an adapted environment, increase in variation of the expressed phenotype in the novel environments. The model should also incorporate the role of the environment as an inducer. While evolution of mutational robustness of GRNs is very well documented \citep{wagner1996does, nagata2020emergence, kaneko2022evolution}, none have explicitly discussed its role in facilitating adaptation to sudden large environmental changes. 

A notable exception is \citet{espinosa2011phenotypicrobustness} who investigated the role of mutational robustness in facilitating variability in response to environmental perturbations. They represented environmental cues as initial conditions. Hence, a large change in initial conditions corresponds to a large environmental change. They observed that populations that exhibit high mutational robustness harbour a larger variation of genotypes and express a wider variation in phenotypes after a large environmental perturbation. This shows that the Wagner model is sufficient for describing uncovering of cryptic genetic variations in response to large environmental changes.

\subsection{Change in regulation or form}


The next criterion for validating plasticity-led evolution is evidence of a genetic change in the regulation or form of the phenotype. \citet{levis2016evaluating} proposed measuring changes to the reaction norm to experimentally validate this criterion. However, this approach is inappropriate for highly nonlinear developmental systems such as the Wagner model. In computational models, we can instead measure the degree of plasticity directly as the sensitivity of the phenotype with respect to a change in the environmental cue.




A computational model examining the change in plasticity should express a change in mutational and environmental sensitivity (or robustness) over evolution. The acquisition of mutational robustness as a by-product of increased fitness is well documented \citep{wagner1996does, espinosa2011phenotypicrobustness, espinosa2016selection, kaneko2022evolution}. However, few works have explicitly investigated the link between mutational and environmental robustness.

An exception is \citet{espinosa2011phenotypicplasticity}, who noted that mutations and small environmental perturbations produce a similar range of phenotypes, suggesting that environmental robustness is correlated to mutational robustness. By investigating the genotype space in terms of expressed phenotypes, they also validated that this change in robustness is reflected in mutations in the genome. This is in fact genetic assimilation of novel adaptive traits introduced by \citet{waddington1953genetic}: reduced response to (small) environmental changes as a result of adaptation.


However, genetic assimilation of a single phenotype is not necessarily inevitable in the Wagner model. In fact, \citet{nagata2020emergence} and \citet{kaneko2022evolution} observed the emergence of bistability, which we interpret as a form of increased phenotypic plasticity, in GRNs. { Moreover, \citet{fierst2011history} and \citet{espinosa2021recombination} evolved populations that exhibit increased phenotypic plasticity by selecting individuals that express a different phenotype when input a different environmental cue.} Therefore, the Wagner model is sufficient for describing the increase or decrease to the level of phenotypic plasticity in different contexts.





\subsection{Evidence of adaptive refinement}

Finally, populations exposed to selection under a novel environment should exhibit refinement in adaptive traits. This condition follows from two assumptions inspired by \citet{west2003developmental}\citep{levis2016evaluating, levis2019plasticity}: populations exposed to selection under the novel environment should produce the novel trait more frequently than populations that are not exposed to said novel environment and traits in a population in which it is expressed (and exposed to selection) more frequently should evolve greater and more rapid refinement. As a corollary, individuals of such populations should express higher levels of fitness to the novel environment than lineages that were never exposed to selection under the novel environment \citep{levis2016evaluating, levis2019plasticity}. 


Experimental validation of this criterion requires identification of lineages exhibiting plasticity between environments as a proxy for the ancestral population state \citep{levis2019plasticity}. Computationally, this is not required because one can directly track the population over evolution. This is one of the key advantages of computational modelling which can complement physical experiments. However, the Wagner model, being an evolutionary algorithm itself, continuously tries to optimize the phenotype through exploration of mutations, and hence trivially exhibits adaptive refinement.


Nevertheless, one may attempt to recreate a computational analogy of validation in sensu \citet{levis2019plasticity} by comparing two populations trained to fit to an ancestral environment and then bringing one of the populations to fit to a novel environment while the other continues to train in the ancestral environment. While there are no explicit computational studies that perform such an experiment, we may synthesize some predictions by reviewing results from a collection of computational studies. \citet{espinosa2011phenotypicrobustness} noted that the response upon such a large environmental change on the ancestral-proxy lineage, which is already adapted to the ancestral environment, does not change after further selection under the ancestral environment. On the other hand, traits become more adaptive, robust \citep{wagner1996does} and frequently expressed \citep{espinosa2011phenotypicplasticity} after selection in the novel environment. This validates that the Wagner model can describe adaptive refinement under the novel environment.

\section{Plasticity-led evolution and learning theory}

We remark that the Wagner model is mathematically equivalent to a recurrent neural network (RNN) \citep{watson2016can}. Numerous authors have pointed out analogies between evolution and learning \citep{watson2014evolution,  kouvaris2017evolution, szilagyi2020phenotypes} which shed new light on plasticity-led evolution. RNNs process input data to give predictions as outputs. In development, GRNs process environmental cues to express phenotypes as outputs. However, the learning process itself differs between RNNs and GRNs. RNNs are trained by changing their network parameters in order to minimize the prediction error, which bears similarity to Lamarckian evolution. On the other hand, GRNs are selected based on their phenotypes, then mutations are randomly introduced to their network parameters. (Note the absence of direct causality between selection and mutation, a requirement for Darwinian evolution.)  Let us further examine these analogies in the context of plasticity-led evolution.

Generalization is defined as the capability of a RNN to produce good predictions for inputs not in the training set \citep{bishop2006pattern}. By treating the output of a RNN outside the training set to be analogous to plastic response of an individual in a novel environment, we can regard adaptive plastic response as a form of generalization. {\citet{parter2008facilitated} demonstrated that Boolean network model of evolution, which is a variant of the Wagner model, can learn from past environments to generalize to new environments. \citet{watson2014evolution} also showed the generalization capabilities of GRNs to produce (potentially) adaptive phenotypes to fit future environments.} \citet{kouvaris2017evolution} further analogized generalization to facilitated variation \citep{gerhart2007theory} which they considered a prerequisite for evolvability. {These results suggest that adaptive plastic response may be a universal feature of GRNs.}




We remark that there is a lack of analogy for cryptic genetic variation in machine learning. This is because ``neutral'' weights that do not contribute to increased performance tend to be eliminated to simplify the models and to increase prediction robustness. In contrast, in plasticity-led evolution, robustness breaks down when a population is exposed to a large environmental change. This breakdown uncovers cryptic genetic mutations, which in turn increases phenotypic variation and hence evolvability. We note that a kind of phase transition of robustness should exist in response to environmental changes, which may be an interesting topic for further research.

We would also like to highlight an analogy between retraining of RNNs and genetic accommodation of novel phenotypes. Retraining RNNs changes the network architecture and fine-tunes existing weights to increase performance on new data \citep{bishop2006pattern}. Analogously, genetic accommodation results in changes to regulatory mechanisms of the novel phenotype as well as adaptive refinement of traits through accumulation of adaptive mutations. In machine learning, RNNs are retrained when their outputs no longer accurately reflect new data \citep{bishop2006pattern}. In plasticity-led evolution, genetic accommodation occurs when the constitutive phenotype does not fit the novel environment.

\begin{figure}
  \caption{\label{box: dir}Possible future directions for computational modelling of plasticity-led evolution}
  \hrule
\begin{itemize}
\item Initial adaptive plastic response as a universal property of the GRNs and developmental process.
\item Identify key universal signatures characterizing Plasticity-Led Evolution. These signatures must discriminate between Mutation-Led and Plasticity-Led Evolution.
\item Correlation between environment-as-inducer and environment-as-selector to properly study adaptiveness of plastic response.
\item Breakdown points of robustness.
\item Effects of hierarchical network structures on Plasticity-Led Evolution. This should include effects of epigenetics and higher-order regulatory elements.
\end{itemize}
  \hrule
\end{figure}

\section{Conclusion}

We highlighted the role of phenotypic plasticity in facilitating adaptation and survival in the face of a large environmental change. This led to the discussion on how phenotypic plasticity can change over evolution. Taken together, this motivates an alternative form of evolution called plasticity-led evolution.  We reviewed the results on the GRN model introduced by \citet{wagner1996does} and its variants. We propose that the Wagner model forms a minimal model for computationally investigating plasticity-led evolution. We then looked into the extent to which notions arising from machine learning theory can describe plasticity-led evolution. We noted several analogies between learning theory and plasticity-led evolution as well as discrepancies arising from motive and implementation of machine learning. We would like to highlight that deep neural networks are analogous to the hierarchical structure of biological networks including epigenome, protein-protein interactions and other higher order biological processes. Studying deep neural networks can provide insights into how hierarchical network structures can enhance evolvability. We summarized future directions in Box \ref{box: dir}.

\section*{Acknowledgments}
The authors thank David Marshall and Daphne Teck Ching Lai for helpful comments, and Haziq Jamil for stimulating discussion and reading the manuscript.



\end{document}